\documentclass[pre,twocolumn,superscriptaddress]{revtex4-1}  %floatfix

\usepackage[paperwidth=210mm,paperheight=297mm,centering,hmargin=2cm,vmargin=2.5cm]{geometry}

\usepackage[pdftex]{graphicx}
\usepackage{amsmath,amsfonts,amssymb}
\usepackage{morefloats}
\usepackage{slashbox}
\usepackage{natbib}

\usepackage{hyperref}
\usepackage{xcolor} 

\definecolor{bluemoi}{rgb}{0.25,0.50 ,0.75} 

\hypersetup{
  bookmarksopen=false,
  pdftitle=" ",
  pdfauthor=" ", 
  pdfsubject=" ", 
  pdftoolbar=true, % toolbar hidden
  pdfmenubar=true, %menubar shown
  pdfhighlight=/O, %effect of clicking on a link
  colorlinks=true, %couleurs sur les liens hypertextes
  pdfpagemode=UseNone, %aucun mode de page
  pdfpagelayout=SinglePage, %ouverture en simple page
  pdffitwindow=true, %pages ouvertes entierement dans toute la fenetre
  linkcolor=bluemoi, %couleur des liens hypertextes internes
  citecolor=bluemoi, %couleur des liens pour les citations
  urlcolor=bluemoi %couleur des liens pour les url
}

\makeatletter
\renewcommand{\figurename}{\sf \textbf{Figure}}
\renewcommand{\thefigure}{\arabic{figure}}
\renewcommand{\fnum@figure}{\sf\textbf{\figurename}~\textbf{\thefigure}}
\renewcommand{\tablename}{\sf\textbf{Table}}
\renewcommand{\thetable}{\arabic{table}}
\renewcommand{\fnum@table}{\sf\textbf{\tablename}~\textbf{\thetable}}
\makeatother

\begin{document}

\title{Touristic site attractiveness seen through Twitter}

\author{Aleix Bassolas}\affiliation{Instituto de F\'isica Interdisciplinar y Sistemas Complejos IFISC (CSIC-UIB), Campus UIB, 07122 Palma de Mallorca, Spain}
\author{Maxime Lenormand}\affiliation{Instituto de F\'isica Interdisciplinar y Sistemas Complejos IFISC (CSIC-UIB), Campus UIB, 07122 Palma de Mallorca, Spain}
\author{Ant{\`o}nia Tugores}\affiliation{Instituto de F\'isica Interdisciplinar y Sistemas Complejos IFISC (CSIC-UIB), Campus UIB, 07122 Palma de Mallorca, Spain}
\author{Bruno Gon{\c c}alves}\affiliation{Center for Data Science, New York University, 726 Broadway, 7th Floor, 10003 New York, USA and Aix Marseille Universit{\' e}, Universit{\'e } de Toulon, CNRS, CPT, UMR 7332, 13288 Marseille, France}
\author{Jos\'e J. Ramasco}\affiliation{Instituto de F\'isica Interdisciplinar y Sistemas Complejos IFISC (CSIC-UIB), Campus UIB, 07122 Palma de Mallorca, Spain}

\begin{abstract} 
Tourism is becoming a significant contributor to medium and long range travels in an increasingly globalized world. Leisure traveling has an important impact on the local and global economy as well as on the environment. The study of touristic trips is thus raising a considerable interest. In this work, we apply a method to assess the attractiveness of $20$ of the most popular touristic sites worldwide using geolocated tweets as a proxy for human mobility. We first rank the touristic sites based on the spatial distribution of the visitors' place of residence. The Taj Mahal, the Pisa Tower and the Eiffel Tower appear consistently in the top 5 in these rankings. We then pass to a coarser scale and classify the travelers by country of residence. Touristic site's visiting figures are then studied by country of residence showing that the Eiffel Tower, Times Square and the London Tower welcome the majority
of the visitors of each country. Finally, we build a network linking sites whenever a user has been detected in more than one site. This allow us to unveil relations between touristic sites and find which ones are more tightly interconnected.    
\end{abstract}

\maketitle

\section*{INTRODUCTION}

Traveling is getting more accessible in the present era of progressive globalization. It has never been easier to travel, resulting in a significant increase of the volume of leisure trips and tourists around the world (see, for instance, the statistics of the last UNWTO reports \cite{unwto}). 
Over the last fifty years, this increasing importance of the economic, social and environmental impact of tourism on a region and its residents has led to a considerable number of studies in the so-called geography of tourism \cite{Christaller1964}. In particular, geographers and economists have attempted to understand the contribution of tourism to global and regional economy \cite{Williams1991, Hazari1995, Durbarry2004, Proenca2008, Matias2009} and to assess the impact of tourism on local people \cite{Long1990, Madrigal1993, Jurowski1997, Lindberg1997, Andereck2000, Gursoy2002}. 

These researches on tourism have traditionally relied on surveys and economic datasets, generally composed of small samples with a low spatio-temporal resolution. However, with the increasing availability of large databases generated by the use of geolocated information and communication technologies (ICT) devices such as mobile phones, credit or transport cards, the situation is now changing. Indeed, this flow of information has notably allowed researchers to study human mobility patterns at an unprecedented scale \cite{Brockmann2006, Gonzalez2008, Song2010, Noulas2012, Hawelka2014, Lenormand2015a}. In addition, once these data are recorded, they can be aggregated in order to analyze the city's spatial structure and function \cite{Reades2007, Soto2011, Frias2012, Pei2014, Louail2014, Grauwin2014, Lenormand2015c, Louail2015} and they have also been successfully tested against more traditional data sources \cite{Lenormand2014a, Deville2014, Tizzoni2014}. In the field of tourism geography, these new data sources have offered the possibility to study tourism behavior at a very high spatio-temporal resolution \cite{Asakura2007, Shoval2007, Girardin2008, Freytag2010, Poletto2012, Poletto2013, Hawelka2014, Bajardi2015}.

In this work, we propose a ranking of touristic sites worldwide based on their attractiveness measured with geolocated data as a proxy for human mobility. Many different rankings of most visited touristic sites exist but they are often based on the number of visitors, which does not really tell us much about their attractiveness at a global scale. Here we apply an alternative method proposed in \cite{Lenormand2015b} to measure the influence of cities. The purpose of this method is to analyze the influence and the attractiveness of a site based on the average radius traveled and the area covered by individuals visiting this site. More specifically, we select $20$ out of the most popular touristic sites of the world and analyze their attractiveness using a dataset containing about $10$ million geolocated tweets, which have already demonstrated their efficiency as useful source of data to study mobility at a world scale \cite{Hawelka2014, Lenormand2015b}. In particular, we propose three rankings of the touristic sites' attractiveness based on the spatial distribution of the visitors' place of residence, we show that the Taj Mahal, the Pisa Tower and the Eiffel Tower appear always in the top 5. Then, we study the touristic site's visiting figures by country of residence, demonstrating that the Eiffel Tower, Times Square and the London Tower attract the majority of the visitors. To close the analysis, we focus on users detected in more than one site and explore the relationships between the $20$ touristic sites by building a network of undirected trips between them.

\begin{figure*}
   \includegraphics[width=\textwidth]{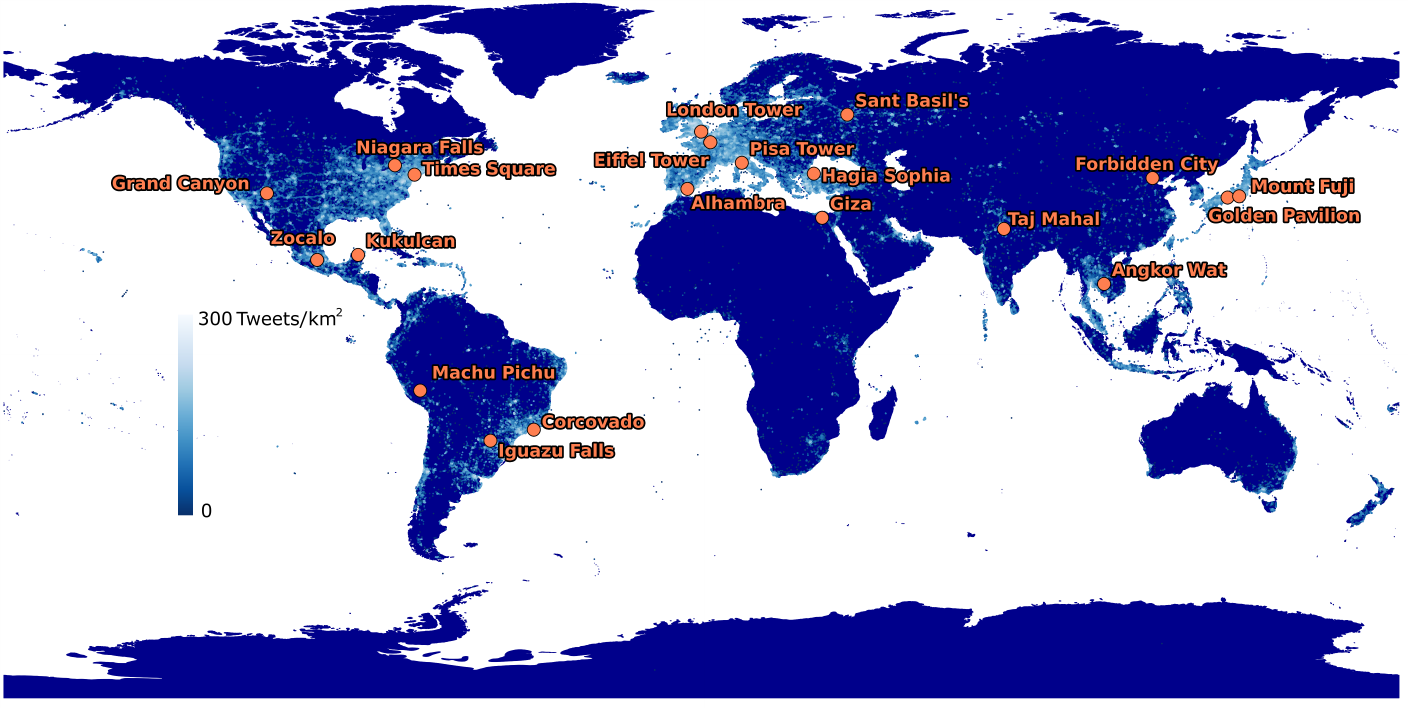}
   \caption{\textbf{Density of geolocated tweets and positions of the touristic sites.}\label{Fig1}}
\end{figure*}

\section*{MATERIALS AND METHODS}

The purpose of this study is to measure the attractiveness of $20$ touristic sites taking into account the spatial distribution of their visitors' places of residence. To do so, we analyze a database containing $9.6$ million geolocated tweets worldwide posted in the period between September 10, 2010 and October 21, 2015. The dataset was built by selecting the geolocated tweets sent from the touristic places in the general streaming and requesting the time-lines of the users posting them. The touristic sites boundaries have been identified manually.  Collective accounts and user exhibiting non-human behaviors have been removed from the data by identifying users tweeting too quickly from the same place, with more than 9 tweets during the same minute and from places separated in time and space by a distance larger than what is possible to be covered by a commercial flight (with an average speed of 750 km/h). Their spatial distributions and that of the touristic sites can be seen in Figure \ref{Fig1}.

In order to measure the site attractiveness, we need to identify the place of residence of every user who have been at least once in one of the touristic sites. First, we discretize the space by dividing the world into squares of equal area ($100 \times 100 \text{ km}^2$) using a cylindrical equal-area projection. Then, we identify the place most frequented by a user as the cell from which he or she has spent most of his/her time. To ensure that this most frequented location is the actual user's place of residence the constraint that at least one third of the tweets has been posted from this location is imposed. The resulting dataset contains about $59,000$ users' places of residence. The number of valid users is shown in Table \ref{Tab1} for each touristic site. In the same way, we identify the country of residence of every user who have posted a tweets from one of the touristic sites during the time period.

\begin{table}
\begin{center}
\small
    \begin{tabular}{ | l | l |}
    \hline
\textbf{Site}  &  \textbf{Users}\\  
\hline  
Alhambra (Granada, Spain)  &  1,208  \\  \hline   
Angkor Wat (Cambodia)   &    947  \\  \hline
 Corcovado (Rio, Brazil)  & 1,708 \\  \hline   
Eiffel Tower (Paris, France) & 11,613 \\  \hline
Forbidden City (Beijing, China)  &  457 \\  \hline    
Giza  (Egypt) &    205  \\  \hline
Golden Pavilion (Kyoto, Japan)   &  1,114  \\  \hline    
Grand Canyon (US)  &    1,451  \\  \hline
Hagia Sophia (Istanbul, Turkey)  &  2,701  \\  \hline    
Iguazu Falls (Argentina-Brazil)  &    583  \\  \hline
Kukulcan (Chichen Itz\'a, Mexico)  &  209  \\  \hline    
London Tower (London, UK)   &    3,361  \\  \hline
Machu Pichu (Peru)  &  987  \\  \hline    
Mount Fuji (Japan)  &    2,241  \\  \hline
Niagara Falls (Canada-US)  &  920  \\  \hline  
Pisa Tower (Pisa, Italy) & 1,270  \\  \hline
Saint Basil's (Moscow, Russia)  &    262  \\  \hline
Taj Mahal (Agra, India) &  378      \\  \hline
Times Square (NY, US)  &    13,356 	     \\  \hline    
Zocalo (Mexico City, Mexico)  &    16,193  \\  \hline
\end{tabular}
\end{center}
\caption{\textbf{Number of valid Twitter users identified in each touristic sites.}}
\label{Tab1}
\end{table} 

\begin{figure*}
\begin{center}
\includegraphics[width=\textwidth]{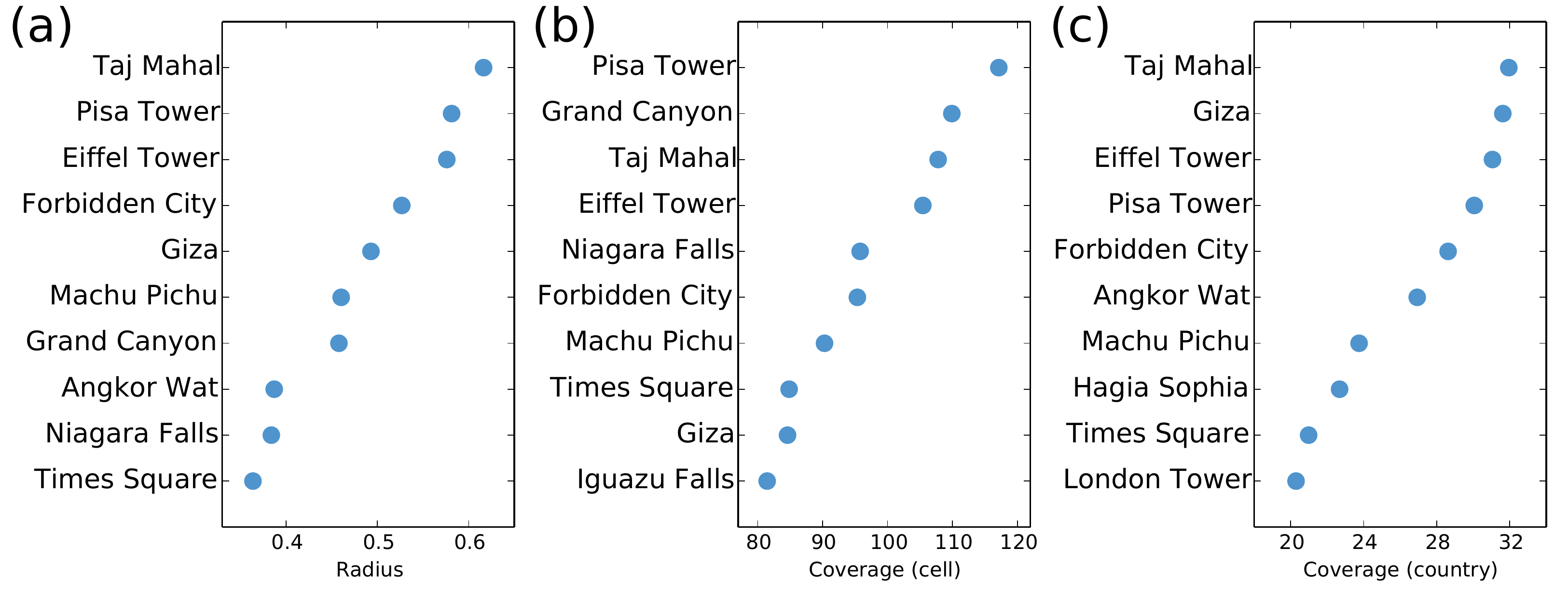}
\caption{\textbf{Ranking of the touristic sites according to the radius and the coverage.} (a) Radius. (b) Coverage (cell). (c) Coverage (country). \label{Fig2}}
\end{center}
\end{figure*}

Two metrics have been considered to measure the attractiveness of a touristic site based on the spatial distribution of the places of residence of users who have visited this site:
\begin{itemize}
\item \textbf{Radius:} The average distance between the places of residence and the touristic site. The distances are computed using the Haversine formula between the latitude and longitude coordinates of the centroids of the cells of residence and the centroid of the touristic site. In order not to penalize isolated touristic sites, the distances have been normalized by the average distance of all the Twitter users' places of residence to the site. It has been checked that the results are consistent if the median is used instead of the average radius for the rankings.
\item \textbf{Coverage:} The area covered by the users' places of residence computed as the number of distinct cells (or countries) of residence.
\end{itemize}

To fairly compare the different touristic sites which may have different number of visitors, the two metrics are computed with $200$ users' place of residence selected at random and averaged over $100$ independent extractions. Note that unlike the coverage the radius does not depend on the sample size but, to be consistent, we decided to use the same sampling procedure for both indicators. 

\section*{RESULTS}

\subsection*{Touristic sites' attractiveness}

We start by analyzing the spatial distribution of the users' place of residence to assess the attractiveness of the $20$ touristic sites. In Figure \ref{Fig2}a and Figure \ref{Fig2}b, the touristic sites are ranked according to the radius of attraction based on the distance traveled by the users from their cell of residence to the touristic site and the area covered by the users' cells of residence. In both cases, the results are averaged over $100$ random selection of $200$ users. The robustness of the results have been assessed with different sample sizes (50, 100 and 150 users), we obtained globally the same rankings for the two metrics. Both measures are very correlated and for most of the site the absolute difference between the two rankings is lower or equal than $2$ positions. However, since the metrics are sensitive to slightly different information both rankings also display some dissimilarities. For example, the Grand Canyon and the Niagara Falls exhibit a high coverage due to a large number of visitors from many distinct places in the US but a low radius of attraction at the global scale.

\begin{figure}[h!]
\begin{center}
\includegraphics[width=8.5cm]{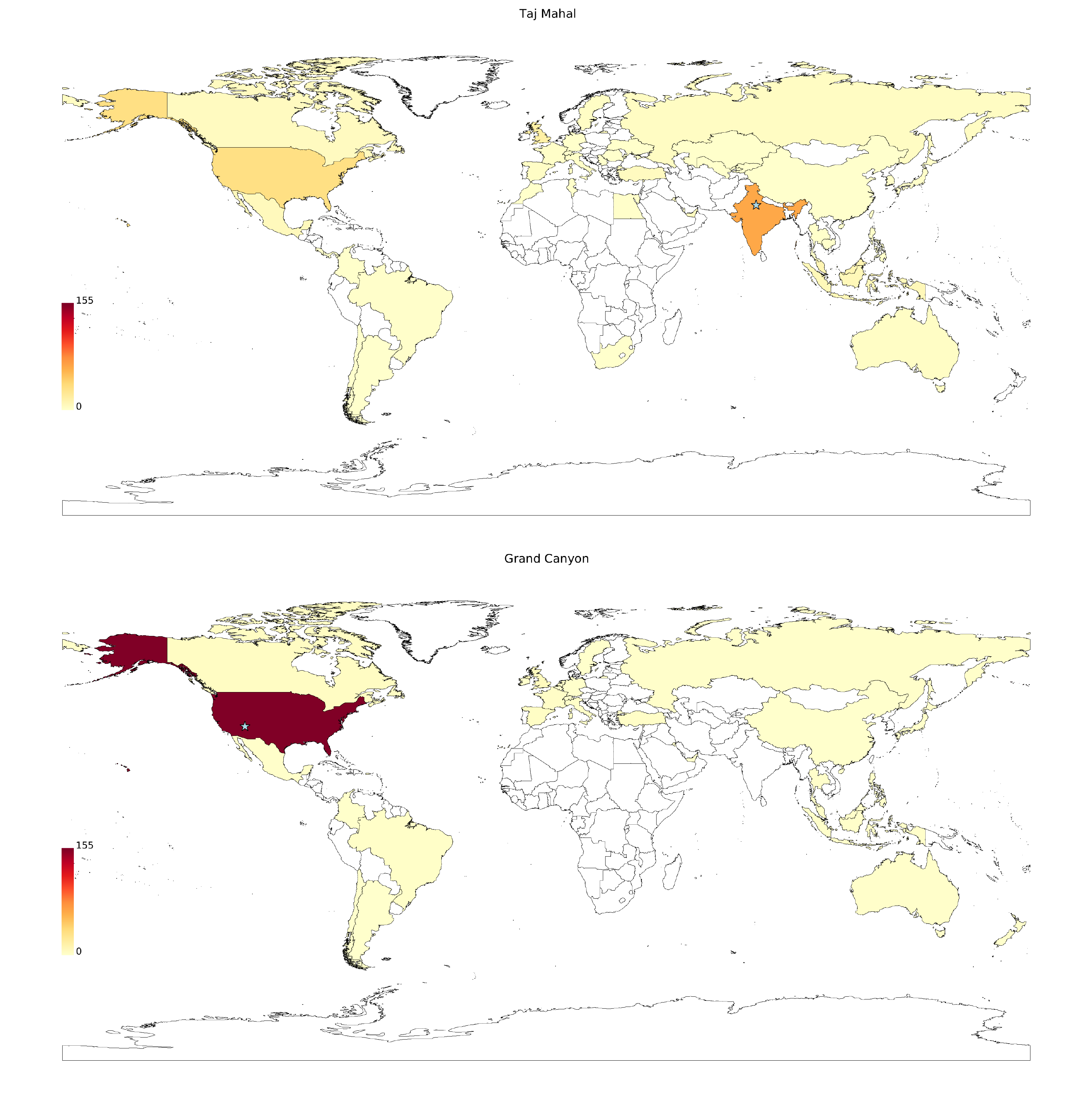}
\caption{\textbf{Heat map of the spatial distribution of the visitors' country of origin for the Taj Mahal and the Grand Canyon. The results have been averaged over $100$ extractions of $200$ randomly selected users.} \label{Fig3}}
\end{center}
\end{figure}

To complete the previous results, we also consider the number of countries of origin averaged over $100$ random selection of $200$ users. This gives us new insights on the origin of the visitors. For example, as it can be observed in Figure \ref{Fig3}, the visitors of the Grand Canyon are mainly coming from the US, whereas in the case of the Taj Mahal the visitors' country of residence are more uniformly distributed. Also, it is interesting to note that in most of the cases the nationals are the main source of visitors except for Angkor Wat (Table \ref{Tab2}). Some touristic sites have a national attractiveness, such as the Mont Fuji or Zocalo hosting about 84\% and 93\% of locals, whereas others have a more global attractiveness, this is the case of the Pisa Tower and the Machu Pichu welcoming only 21\% of local visitors. 

\begin{table*}
\begin{center}
\small
\begin{tabular}{ | l | l | l | l | l | l | }
    \hline
\textbf{Site}  &  \textbf{Top 1} & \textbf{Top 2}  & \textbf{Top 3}\\  \hline  
Alhambra (Spain)  &    Spain   71.14$\%$   &   US   6.06$\%$   &   UK   2.61$\%$  \\  \hline
Angkor Wat (Cambodia)  &    Malaysia   19.64$\%$   &   Philippines   17.4$\%$   &   US   9.59$\%$  \\  \hline
Corcovado (Brazil)  &    Brazil   81.13$\%$   &   US   4.92$\%$   &   Chile   3.08$\%$  \\  \hline
Eiffel Tower (France)  &    France   26.75$\%$   &   US   16.62$\%$   &   UK   8.92$\%$  \\  \hline
Forbidden City (China)  &    China   26.48$\%$   &   US   14.46$\%$   &   Malaysia   10.95$\%$  \\  \hline
Giza (Egypt)  &    Egypt   30.65$\%$   &   US   9.8$\%$   &   Kuwait   5.85$\%$  \\  \hline
Golden Pavilion  (Japan)  &    Japan   60.74$\%$   &   Thailand   11.72$\%$   &   US   4.84$\%$  \\  \hline
Grand Canyon (US)  &    US   75.79$\%$   &   UK   2.87$\%$   &   Spain   2.16$\%$  \\  \hline
Hagia Sophia (Turkey)  &    Turkey   71.26$\%$   &   US   5.48$\%$   &   Malaysia   1.67$\%$  \\  \hline
Iguazu Falls (Arg-Brazil-Para)  &    Argentina   48.26$\%$   &   Brazil   26.61$\%$   &   Paraguay   8.63$\%$  \\  \hline
Kukulcan (Mexico)  &    Mexico   73.78$\%$   &   US   10.07$\%$   &   Spain   2.83$\%$  \\  \hline
London Tower (UK)  &    UK   65.61$\%$   &   US   10.24$\%$   &   Spain   2.77$\%$  \\  \hline
Machu Pichu  &    Peru   20.43$\%$   &   US   19.95$\%$   &   Chile   10.43$\%$  \\  \hline
Mount Fuji (Japan)  &    Japan   84.01$\%$   &   Thailand   5.83$\%$   &   Malaysia   2.66$\%$  \\  \hline
Niagara Falls (Canada-US)  &    US   60.5$\%$   &   Canada   16.31$\%$   &   Turkey   3.25$\%$  \\  \hline
Pisa Tower (Italy)  &    Italy   20.85$\%$   &   US   13.56$\%$   &   Turkey   10.95$\%$  \\  \hline
Sant Basil (Russia)  &    Russia   66.71$\%$   &   US   5.06$\%$   &   Turkey   3.77$\%$  \\  \hline
Taj Mahal (India)  &    India   27.97$\%$   &   US   15.59$\%$   &   UK   7.61$\%$  \\  \hline
Times Square (US)  &    US   74.32$\%$   &   Brazil   3.26$\%$   &   UK   2.31$\%$  \\  \hline
Zocalo (Mexico)  &    Mexico   92.22$\%$   &   US   3.1$\%$   &   Colombia   0.77$\%$  \\  \hline
\end{tabular}
\end{center}
\caption{\textbf{The three countries hosting most of the visitors for each touristic site.} The countries are ranked by percentage of visitors.}
\label{Tab2}
\end{table*}

More generally, we plot in Figure \ref{Fig2}c the ranking of touristic sites based on the country coverage. The results obtained are very different than the ones based on the cell coverage (Figure \ref{Fig2}b). Indeed, some touristic sites can have a low cell coverage but with residence cells located in many different countries, this is the case of the Pyramids of Giza, which went up $7$ places and appears now in second position. On the contrary, other touristic sites have a high cell coverage but with many cells in the same country, as in the previously mentioned cases of the Grand Canyon and the Niagara Falls. Finally, the ranks of the Taj Mahal, the Pisa Tower and Eiffel Tower are consistent with the two previous rankings, these three sites are always in the top $5$. Finally, we compare quantitatively the rankings with the Kendall's $\tau$ correlation coefficient which is a measure of association between two measured quantities based on the rank. In agreement with the qualitative observations, we obtain significant correlation coefficients comprised between 0.66 and 0.77 confirming the consistency between rankings obtained with the different metrics.

\subsection*{Touristic site's visiting figures by country of residence}

\begin{figure*}
\begin{center}
\includegraphics[width=14cm]{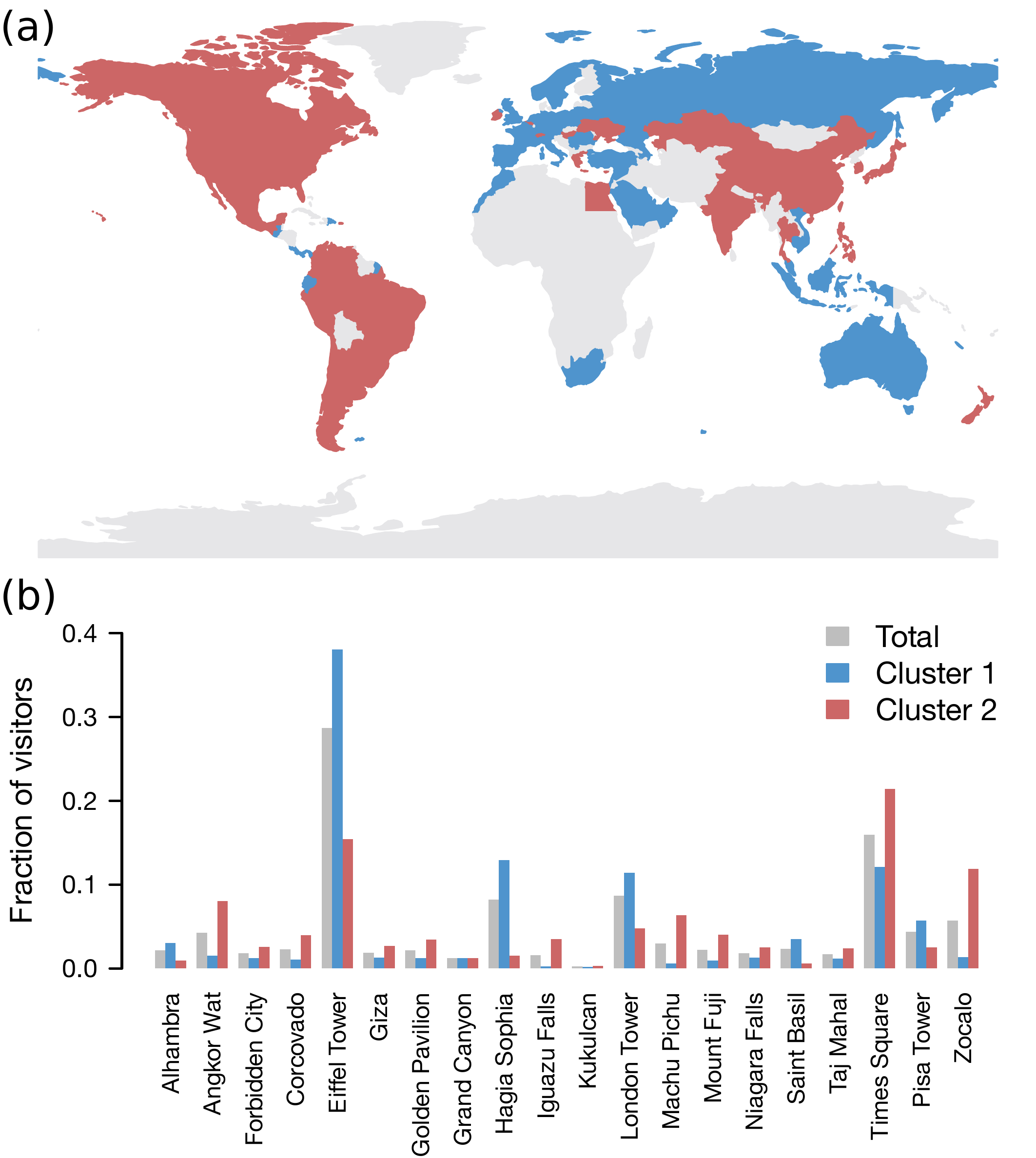}
\caption{\textbf{Clustering analysis.} (a) Map of the spatial distribution of the country of residence according to the cluster.  (b) Fraction of visitors according to the touristic site. \label{Fig4}}
\end{center}
\end{figure*}

We can also do the opposite by studying the touristic preferences of the residents of each country. We extract the distribution of the number of visitors from each country to the touristic sites and normalize by the total number of visitors in order to obtain a probability distribution to visit a touristic site according to the country of origin. This distribution can be averaged over the $70$ countries with the higher number of residents in our database (gray bars in Figure \ref{Fig4}). The Eiffel Tower, Times Square and the London Tower welcome in average $50$\% of the visitors of each country. It is important to note that these most visited touristic sites are not necessarily the ones with the higher attractiveness presented in the previous section. That is the advantage of the method proposed in \cite{Lenormand2015b}, which allows us to measure the influence and the power of attraction of regions of the world with different number of local and non-local visitors.
 
We continue our analysis by performing a hierarchical cluster analysis to group together countries exhibiting similar distribution of the number of visitors according to the touristic sites. Countries are clustered together using the ascending hierarchical clustering method with the average linkage clustering as agglomeration method and the Euclidean distance as similarity metric, respectively. To choose the number of clusters, we used the average silhouette index \cite{Rousseeuw1987}. The results of the clustering analysis are shown in Figure \ref{Fig4}. Two natural clusters emerge from the data, these clusters are without surprise composed of countries which tend to visit in a more significant way touristic sites located in countries belonging to their cluster. The first cluster gather countries of America and Asia whereas the second one is composed of countries from Europe and Oceania.

\subsection*{Network of touristic sites}

\begin{figure*}
\begin{center}
\includegraphics[width=\textwidth]{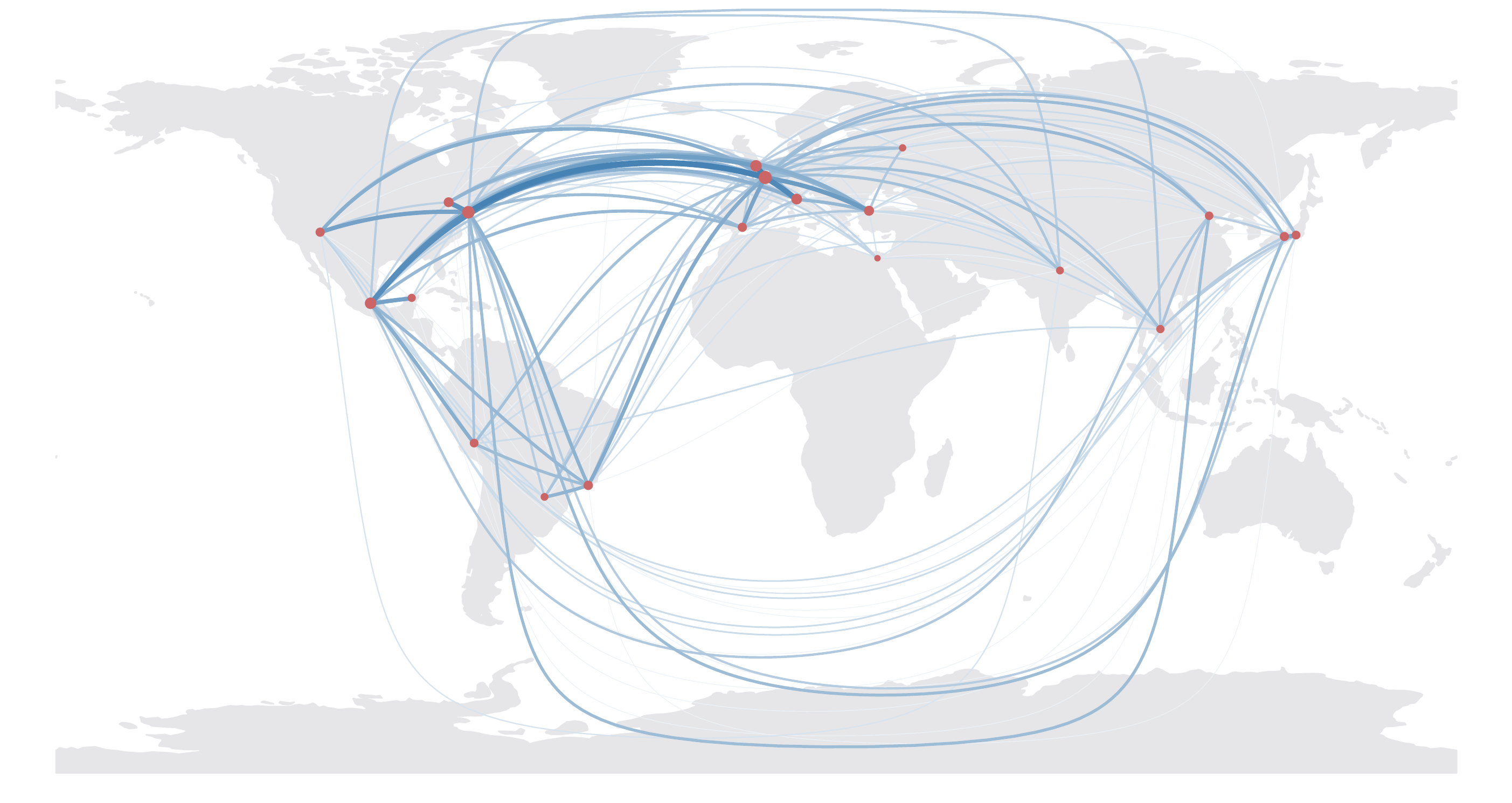}
\caption{\textbf{Network of undirected trips between touristic sites.} The width and the brightness of a link is proportional to its weight. The size of a node is proportional to its weighted degree. \label{Fig5}}
\end{center}
\end{figure*}

In the final part of this work, we investigate the relationships between touristic sites based on the number of Twitter users who visited more than one site during a time window between September 2010 and October 2015. More specifically, we built an undirected spatial network for which every link between two toursitc sites represents at least one user who has visited  both sites. As a co-occurence network, the weight of a link between two sites is equal to the total number of users visiting the connected sites. The network is represented in Figure \ref{Fig5} where the width and the brightness of a link is proportional to its weight and the size of a node is proportional to its weighted degree (strenght). The Eiffel Tower, Times Square, Zocalo and the London Tower appear to be the most central sites playing a key role in the global connectivity of the network (Table \ref{Tab3}). The Eiffel Tower alone accounted for a 25\% of the total weighted degree. The three links exhibiting the highest weights connect the Eiffel Tower with Time Square, the London Tower and the Pisa Tower representing 30\% the total sum of weights. Zocalo is also well connected with the Eiffel Tower and Time Square representing 11\% of the total sum of weights. 
 
\begin{table}
\begin{center}
\small
    \begin{tabular}{ | l | c | }
    \hline
\textbf{Site}  &  \textbf{Node Total Weight}\\  \hline 
Eiffel Tower (France)     &   $0.25$  \\  \hline 
Times Square (US)         &   $0.17$  \\  \hline 
Zocalo (Mexico)           &   $0.10$  \\  \hline 
London Tower (UK)         &   $0.10$  \\  \hline 
Pisa Tower (Italy)        &   $0.06$  \\  \hline 
Hagia Sophia (Turkey)     &   $0.04$  \\  \hline 
Niagara Falls (Canada-US) &   $0.04$  \\  \hline 
Corcovado (Brazil)        &   $0.03$  \\  \hline 
Alhambra (Spain)          &   $0.03$  \\  \hline 
Grand Canyon (US)         &   $0.03$  \\  \hline 
\end{tabular}
\end{center}
\caption{\textbf{Ranking of nodes based on the fraction of the total weight.}}
\label{Tab3}
\end{table}

\section*{DISCUSSION}

We study the global attractiveness of $20$ touristic sites worldwide taking into account the spatial distribution of the place of residence of the visitors as detected from Twitter. Instead of studying the most visited places, the focus of the analysis is set on the sites attracting visitors from most diverse parts of the world. A first ranking of the sites is obtained based on cells of residence of the users at a geographical scale of $100$ by $100$ kilometers. Both the radius of attraction and the coverage of the visitors' origins consistently point toward the Taj Mahal, the Eiffel tower and the Pisa tower as top rankers. When the users' place of residence is scaled up to country level, these sites still appear on the top and we are also able to discover particular cases such as the Grand Canyon and the Niagara Falls that are most visited by users residing in their hosting countries. At country level, the top rankers are the Taj Mahal and the Pyramids of Giza exhibiting a low cell coverage but with residence cells distributed in many different countries. 

Our method to use social media as a proxy to measure human mobility lays the foundation for even more involved analysis. For example, when we cluster the sites by the country of the origin of their visitors, two main clusters emerge: one including the Americas and the Far East and the other with Europe, Oceania and South Africa. The relations between sites have been also investigated by considering users who visited more than one place. An undirected network was built connecting sites visited by the same users. The Eiffel Tower, Times Square, Zocalo and the London Tower are the most central sites of the network. 

In summary, this manuscript serves to illustrate the power of geolocated data to provide world wide information regarding leisure related mobility. The data and the method are completely general and can be applied to a large range of geographical locations, travel purposes and scales. We hope thus that this work contribute toward a more agile and cost-efficient characterization of human mobility. 

\vspace*{0.5cm}
\section*{ACKNOWLEDGEMENTS}

Partial financial support has been received from the Spanish Ministry of Economy (MINECO) and FEDER (EU) under project INTENSE@COSYP (FIS2012-30634),  and from the EU Commission through project INSIGHT. The work of ML has been funded under the PD/004/2013 project, from the Conselleria de Educaci\'on, Cultura y Universidades of the Government of the Balearic Islands and from the European Social Fund through the Balearic Islands ESF operational program for 2013-2017. JJR acknowledges funding from the Ram\'on y Cajal program of MINECO. BG was partially supported by the French ANR project HarMS-flu (ANR-12-MONU-0018).

\bibliographystyle{unsrt}
\bibliography{TouristicSite}

\end{document}